\documentstyle[aps,amssymb,twocolumn,graphicx,psfrag]{revtex}
\DeclareSymbolFont{rsfscript}{OMS}{rsfs}{m}{n}
\DeclareSymbolFontAlphabet{\mathrsfs}{rsfscript}
\renewcommand{\Re}{\mathop{\mathrm{Re}}\nolimits}
\renewcommand{\Im}{\mathop{\mathrm{Im}}\nolimits}
\newcommand{\Real}{{{\mathbb{R}}}}

\newcommand{\Rtwo}{{\Real^2}}

\draft % \draft command makes pacs numbers print
\title{Nodal domains statistics - a criterion for quantum chaos}
\author{Galya Blum, Sven Gnutzmann  and Uzy Smilansky }
\address{The Weizmann Institute of Science, 76100 Rehovot, Israel}

\date{\today}

\begin{document}
\maketitle
\begin{abstract}
  We consider the distribution of the (properly normalized) numbers of
nodal domains of
wave functions in 2-$d$ quantum billiards. We show that these distributions
distinguish
clearly between systems with integrable (separable)  or chaotic underlying
classical
dynamics, and
for each case the limiting distribution is universal (system independent).
Thus, a
new criterion for quantum chaos is provided by the statistics of the wave
functions,
which complements the well established criterion based on spectral statistics.

\end{abstract}
\label{eq: helmholtz}
%05.45.Mt   Semiclassical chaos (``quantum chaos'')
%03.65.Ge   Solutions of wave equations: bound states
%\begin{multicols}{2}
The attempts to discern the ``finger-prints" of the underlying classical
dynamics
in the patterns observed in the quantum wave-functions is one of the main
themes
of quantum chaos \cite{Shnirelman,Gutzwiller,Berry77,Heller,Aurich,Simmel}.
Here,
we present a new approach which is based on the distribution of the number
of the
nodal domains of wave-functions. We show that this distribution makes a clear
distinction between integrable (separable)  and chaotic systems, and we study
the universal features associated with each of the cases. As a convenient
paradigm we
consider ``quantum billiards" in 2-$d$ whose classical dynamics correspond to
geodesic flows  with specular reflections on the domain boundaries.  The
corresponding  Schr\"odinger  (Helmholtz) equation reads:
\begin{equation}
-\bigtriangleup \Psi ({\bf r}) =E \Psi ({\bf r})  \ \ ; \ \ {\bf r} \in
\Omega \ .
\label{eq:helmholtz}
\end{equation}
where  $\Omega$ is a connected compact domain on a 2-$d$
Riemannian manifold, and $\bigtriangleup$ is the corresponding Laplacian.
If $\Omega$
has boundaries, Dirichlet boundary conditions are assumed throughout.

Considering only real solutions of (\ref {eq:helmholtz}), the {\it nodal
lines} are
the zero sets of $\Psi ({\bf r})$. The {\it nodal domains} are the
connected domains
where $\Psi ({\bf r})$ has a constant sign. The distribution of the number
of nodal
domains is the subject of the present note.
Courant \cite{Courant} was the first to address this counting problem.
Ordering the
eigenfunctions by the magnitude of the corresponding eigenvalues, and
denoting by
$\nu_n$ the number of nodal domains of the $n$'th eigenfunction, Courant
proved that
\begin{equation}
  \nu_n \le n \ .
\label{eq:courant}
\end{equation}
Pleijel \cite{pleijel} showed that the equality in (\ref{eq:courant}) holds
only for a finite sequence of eigenfunctions, and that
\begin{equation}
 \overline{\lim_{n\rightarrow \infty}} \ {\nu_n \over  n }\ \le \left (
{2\over j_1}\right
)^2
\approx 0.691\ ,
\label{eq:pleyel}
\end{equation}
where $j_1 $ is the first zero of the Bessel function  $J_0(x)$.
%(See \cite {Don-Fef} for other properties of the nodal sets, not directly
%related to
%the subsequent discussion).

Here, we associate to each wave-function the normalized
nodal-domain number
\begin{equation}
  \xi_n\  =\  {\nu_n\over n }\  \ , \ \ \ \ 0<\xi_n \le 1 \ .
\label{eq:nndom}
\end{equation}
and study their distribution, defined in the
following way.  Consider the spectral interval
$I_g(E) = [E,E+g E]$,  ($g>0$ arbitrary), and denote
by
$N_I$ the number of eigenvalues in $I_g(E)$. The $\xi$ distribution associated
with $I_g(E)$ is
\begin{equation}
 P(\xi,I_g(E))  ={1\over
N_I}\sum_{E_n\in I_g(E)}\delta (\xi - {\nu_n\over n}) \ .
\label{eq:xidis}
\end{equation}
We shall show the existence of the limiting distribution,
\begin{equation}
P(\xi) = \lim_{E\rightarrow \infty}   P(\xi,I_g(E))  \ ,
\label{eq:limitdis}
\end{equation}
and present its universal features.

 For domains with boundaries, we also study the distribution of the number
of nodal intersections with the boundary, denoted by $\tilde \nu_n$. This
is the
number of times the normal derivative at the boundary vanishes.  In this case,
the appropriate normalized parameter is
\begin{equation}
  \eta_n\  =\  {\tilde \nu_n\over\sqrt n }\ ,
\label{eq:nnbound}
\end{equation}
and we define the distribution $\tilde P(\eta,I_g(E))$ and the limiting
distribution
$\tilde P(\eta)$ as in (\ref {eq:xidis}) and (\ref {eq:limitdis}),
respectively.

We shall first discuss integrable systems, and in particular, those generic
integrable systems which are also {\it separable} \cite {Gutzwiller}. This
set of
systems includes  the rectangular  and elliptic domains in $\Rtwo$,
surfaces of revolution \cite {SB} and Liouville surfaces \cite {KMS}. In
the two
latter cases, we shall deal only with surfaces for which
exist {\it global} action-angle variables for the geodesic flows. The relevant
common features to the systems  mentioned above are that (a) The hamiltonian
$H(I_1,I_2)$ is a homogeneous function of degree 2 of the action variables
$I_{1,2}$
(which are expressed in units of $\hbar$). (b) The contour line $ \Gamma $
defined
by $  H(I_1,I_2)=1$ in the positive quadrant of the $(I_1,I_2)$ plane is
monotonic.
 In the semiclassical limit of interest here,
the spectrum is given by the EBK quantization:
$E_{l,m} = H(l+\alpha_1,m+\alpha_2) + {\mathcal{O}}(\sqrt E)$ where $l,m$
are the integer quantum numbers and $\alpha_{1,2}$  are the Maslov indices.
The
normalized  number of nodal domains (\ref {eq:nndom}) is  $ \xi_{l,m} =
\nu_{l,m} /
n_{l,m}$. Seperability implies
$\nu_{l,m}=l\ m +{\mathcal{O}} (1)$,  and $n_{l,m}={\mathcal
{N}}(E_{l,m})$, where
${\mathcal {N}}(E)$ is the spectral counting function.  To leading order,
${\mathcal {N}}(E)$ can be replaced by the first term in the Weyl series,
${\mathcal {N}}(E) = {\mathcal {A}} E (1  + {\mathcal{O}}({1\over \sqrt
E})) $, where ${\mathcal {A}}$ is the area enclosed by  $\Gamma$ and the
$I_1, I_2$
axes. Substituting these approximate values in (\ref {eq:xidis}),  and
converting the discrete sum into an integral, incure a relative error  of order
${\mathcal{O}}({1 \over \sqrt E}) $. Thus,
\begin{eqnarray}
\label{eq:intermediate}
&&P(\xi ;I_g(E)) = \\
&&{1\over g E {\mathcal A} }\int_{ H(I_1,I_2)\in I_g}
{\rm d}I_1  {\rm d} I_2
\delta\left (
\xi-{I_1\ I_2 \over{\mathcal A } H(I_1,I_2) }\right )  +{\mathcal{O}}({1
\over \sqrt
E}) \ . \nonumber
\end{eqnarray}
The homogeneity of the hamiltonian function  leads to an
expression for $P(\xi,I_g(E))$ which is independent of both $E$ and $g$. This
establishes the existence of the limiting distribution (\ref
{eq:limitdis}). Changing the
integration variables
$(I_1,I_2)\rightarrow ({\mathcal E},s)$,
where ${\mathcal E} =H(I_1,I_2)$ and $s$ is the arc-length along the line
$\Gamma$, (\ref {eq:intermediate}) is reduced to
 \begin{equation}
P(\xi) = {1\over {\mathcal A}}\int_{\Gamma} {\rm d}s\  \delta\left (
\xi-{I_1(s)I_2(s)\over {\mathcal A}}\right )  \  .
\label{eq:xidist}
\end{equation}
This form allows a simple geometrical interpretation: $I_1(s)I_2(s)$
is the area of the rectangle generated by a point $(I_1,I_2)$
 on $\Gamma$ and its projections on the axes.  Thus, $P(\xi)$
is the probability distribution of the  areas of rectangles scaled by
${\mathcal A}$, the
area between $\Gamma$ and the axes. However,
$\Gamma$ is monotonic, and therefore the scaled areas are all smaller than
1. Thus,
 $P(\xi)=0$ for $\xi \ge \xi_m$, where $\xi_m$ is the maximum value of the
scaled areas.
The monotonicity of $\Gamma$ ensures the existence of an explicit function for
$\Gamma$: $I_2= I_2(I_1)$, in terms of which the integral (\ref
{eq:xidist}) can be
computed for every value of $\xi\le \xi_m$. The resulting distribution is
\begin{eqnarray}
P(\xi) =\left\{
\begin{array}{l}
\ \ \ \ \ \  0 \ \ \ \ \ \ \ \ \ \ \ \ \ \ \ \ \ \ \ \ \ \ \ \ \ \ \ \ \ \
\ \ \xi \
\ge\ \xi_m \ ,        \\  \\
\left. {1\over 2}\sum {I_2(I_1)-I_2'(I_1)I_1 \over
I_2(I_1)+I_2'(I_1)I_1} \right|_{\xi={I_1 I_2(I_1)\over{\mathcal A}}} \ \
\xi \ < \
\xi_m
\ .
%{\sigma \over 2}\left (\xi_m-\xi\right )^{-{1\over 2}} \ \ \text{ if } \
%\xi\ \
\end{array}
\right.
\label {eq:integp}
\end{eqnarray}
The sum is over all the real values of $I_1$  which satisfy $\xi={I_1
I_2(I_1)\over{\mathcal A}}$. In the vicinity of $\xi_m$, typically two
solutions coalesce, leading to a square root singularity of $P(\xi)$ near
$\xi_m$.
The vanishing of $P(\xi) $ in the interval $(\xi_m,1)$, and the square root
singularity
at $\xi_m$ are the universal features which characterize the nodal domain
distributions
for separable, integrable systems.

The distribution $\tilde P(\eta)$ can be derived in a
similar way, the only difference being that $\tilde \nu_n$ is twice the
value of one
of the action variables $I_1$, say. (The rectangular billiard is the only
exception, where
$\tilde \nu_n = 2(I_1+I_2)$, for $I_{1,2}>1$). The scaling with respect to
$\sqrt n$
is naturaly called for.  Denoting by $I_1^{(m)}$ the intersection of
$\Gamma$ with the
$I_1$ axis,  the maximum value of $\eta$ is $\eta_m= 2 I_1^{(m)}/\sqrt
{\mathcal A}$. Then,
the analogue of (\ref {eq:integp}) reads,
\begin{eqnarray}
\tilde P(\eta) =\left\{
\begin{array}{l}
\ \ \ \ \ \ \ \ \ \  0 \ \ \ \ \ \ \ \ \ \ \ \ \ \ \ \ \ \ \ \ \ \ \ \ \ \
\ \ \ \ \ \
\eta
\ge \eta_m \ ,        \\  \\
\left. {1\over 4 \sqrt {\mathcal A}}(I_2(I_1)-I_2'(I_1)I_1) \right|_{ I_1=
{\eta
\sqrt {\mathcal A}\over 2}} \ \ \eta   <
\eta_m  \ .
\end{array}
\right.
\label {eq:integptild}
\end{eqnarray}
Near $\eta_m$, $\tilde P(\eta)$ reflects the way that $\Gamma$ intersects
the $I_1$ axis,
and it is not universal.
%%%%
\begin{figure*}%
  \begin{center}%
%    \psfrag{nu}{\small $\nu_0$}
%     \psfrag{C}{\hspace*{-1em}\small $C(\nu_0)$}
%     \psfrag{y}{\hspace*{-.5em}\small $\big|\Fourier_t[C]\big|$}
%     \psfrag{t}{\small $t$}
    \includegraphics[width=\linewidth]
    {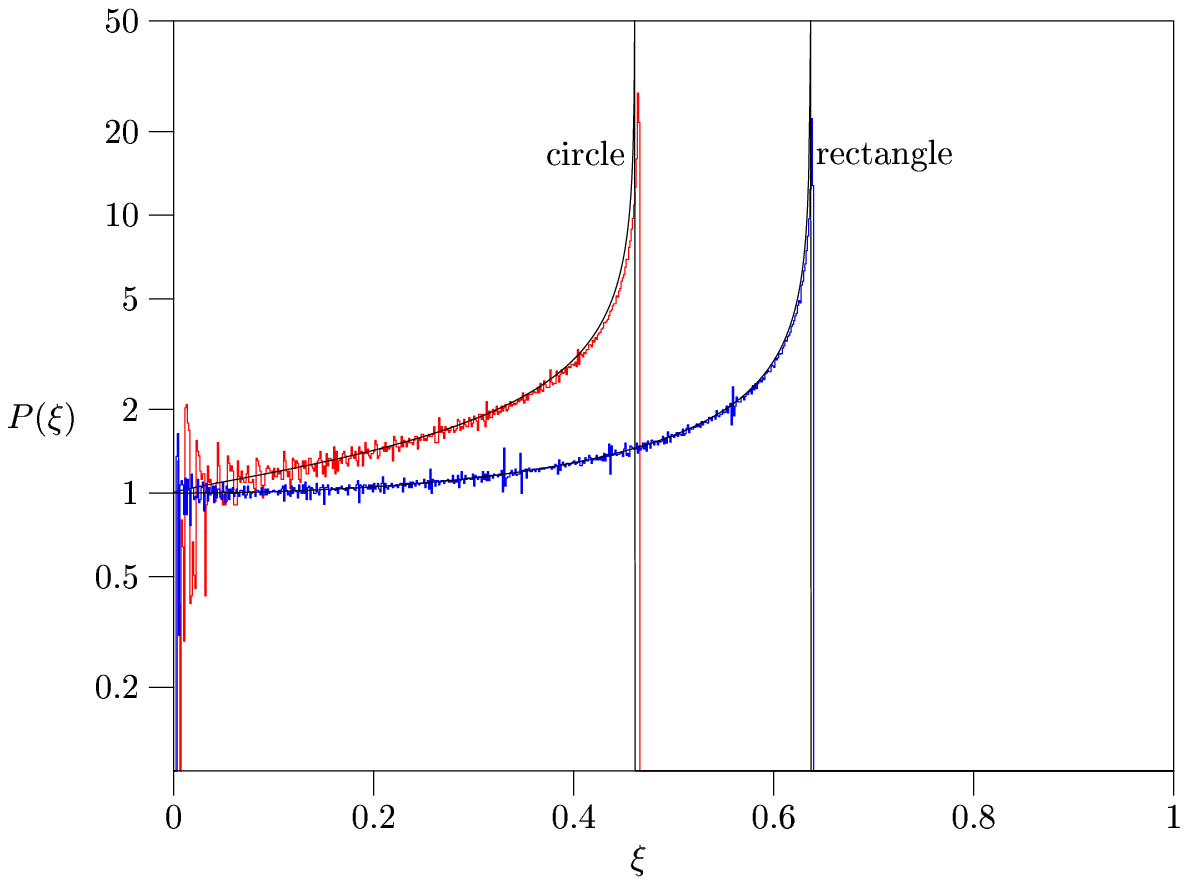}
    \caption{  The nodal domains distributions (\ref {eq:xidist}) for the
rectangular and circular billiards in the spectral intervals:
rectangle- $62500\le n\le 125000$; circle- $30000\le n\le 60000$. Smooth
line- the
limiting distributions (\ref {eq:integp}). }
\label{fig:int.nodal}%
\end{center}%
\end{figure*}%
%%%%
 Numerical simulations for the rectangular and the circular
domains are shown in Figs. (\ref {fig:int.nodal}, \ref
{fig:int.boundnodal}), together
with the predicted limiting distributions. The deviations can be explained
by considering
the next to leading contributions in $1/\sqrt E$. The  singularities
seen in  the numerical data mark the contributions of periodic orbits \cite
{BlumGnUS}.

%%%%
\begin{figure*}%
  \begin{center}%
%    \psfrag{nu}{\small $\nu_0$}
%    \psfrag{C}{\hspace*{-1em}\small $C(\nu_0)$}
%    \psfrag{y}{\hspace*{-.5em}\small $\big|\Fourier_t[C]\big|$}
%    \psfrag{t}{\small $t$}
    \includegraphics[width=\linewidth]
    {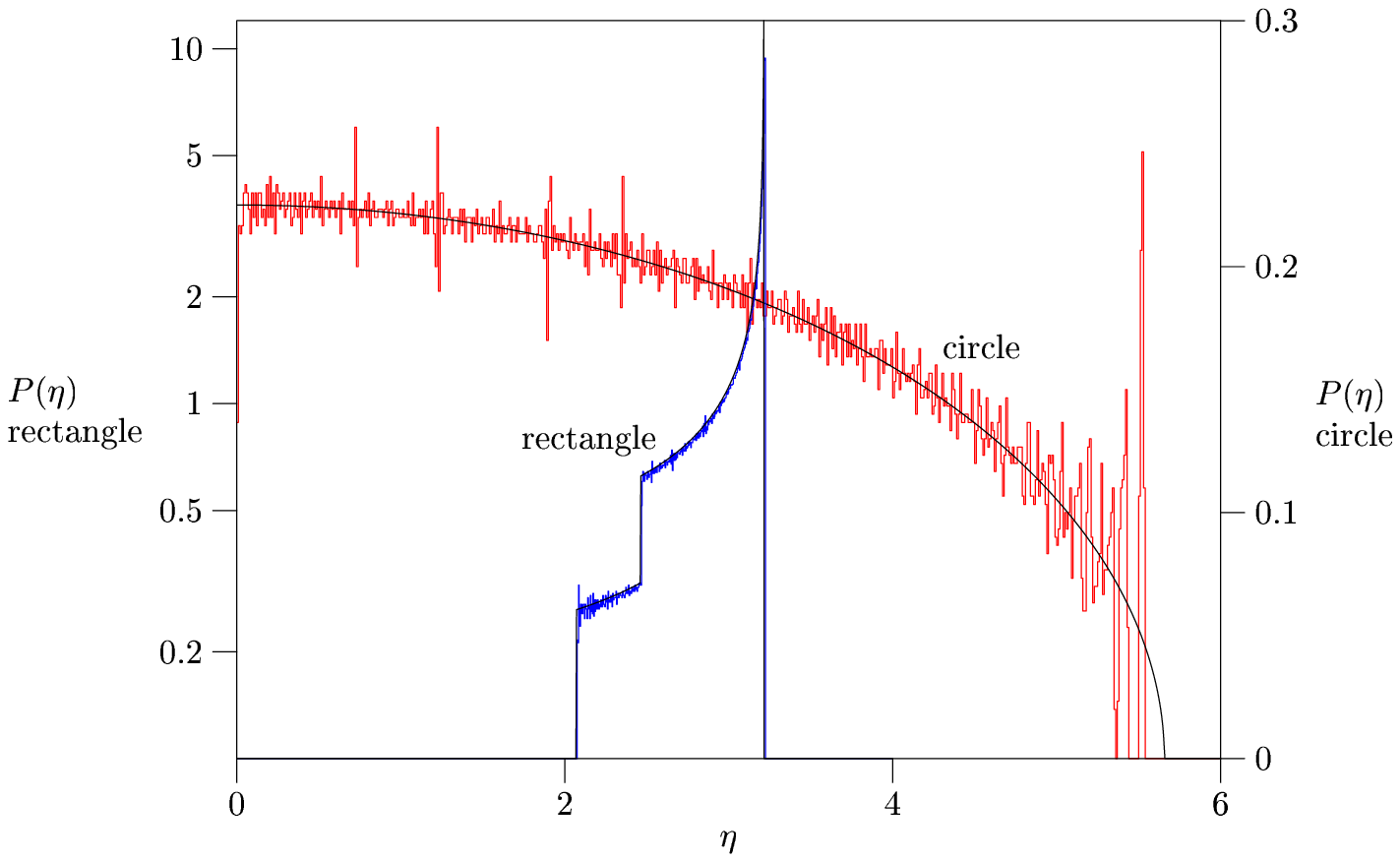}
    \caption{%
     The distribution of the normalized numbers of boundary intersections
for the
rectangular and circular billiards, in the same spectral intervals as in
Fig. 1.
Smooth line- the limiting distributions (\ref {eq:integptild}). }
\label{fig:int.boundnodal}%
\end{center}%
\end{figure*}%
%%%%
The most important feature of wavefunctions corresponding to separable
systems is that the nodal set forms a simple grid of intersecting nodal
lines. In
chaotic systems, nodal intersections may occur, and  when $q$ nodal lines
intersect
at a point, they do so at angles ${\pi\over q}$. The intersections are rare,
and are unstable against small perturbations. There is no known analytical
expression for
the number of nodal domains, and therefore  we cannot offer an expression
for the limiting distribution of the normalized numbers of nodal domains for
chaotic billiards.  However we shall present physical arguments augmented
by numerical
evidence which support the conjecture that the nodal domains of {\it random
waves}
reproduce the observed distributions of chaotic billiards \cite {Berry77}.
While this work
was in progress, Bogomolny and Schmit \cite {bogoschmidt} proposed a method
to compute the
nodal domain distributions for random waves, based on a percolation model.
We shall compare
our numerical result to their predictions.

 The scattering approach to the quantization of billiards \cite
{USleshouches,KlakowUS} implies that the wavefunction can be well
approximated by
\begin{equation}
\Psi_n({\bf r})\approx
\sum_{l=-L}^{L} a_l\ J_l(k_nr)\ {\rm e}^{i l\theta}  \ , \ a_{-l}
=a_l^{*}(-1)^l.
  \label{eq:wavefun}
\end{equation}
Here,  $k=\sqrt E, \  L=\left [ k{\mathcal{L}}/2\pi\right ] $ and
${\mathcal{L}}$ is
the boundary length. The $2L+1$
coefficients $\{a_l\}$ are the  eigenvector of the scattering matrix $S(k=k_n)$
corresponding to an  eigenvalue $1$.  The $S$ matrix is the semiclassical
quantization of the billiard bounce map. For hyperbolic billiard maps,  quantum
ergodicity implies that the vectors $\{a_l\}$ are uniformly distributed on the
sphere. For large $L$,  this distribution can be approximated by considering
$\{ \Re (a_l),\Im (a_l) \}$ as  independent random Gaussian variables with
$\langle |a_l|^2\rangle=1$. Hence, the set of eigenfunctions (\ref
{eq:wavefun})
are conjectured to be simulated by the Gaussian random wave ensemble.

To check the conjectured relation between wave functions of chaotic
billiards and the
ensemble of random waves, we computed the lowest 1637 (1483) eigenfunctions
of the stadium (Sinai) billiards. The  nodal
domains of the billiard wave functions  were counted using the Hoshen-Kopelman
method \cite {HK}.  We performed the following tests.

\begin{figure*} %
 % \begin{center}%
%\vspace{3.0cm}
\hspace{-.9cm}
    \includegraphics[width=.9\linewidth]
    {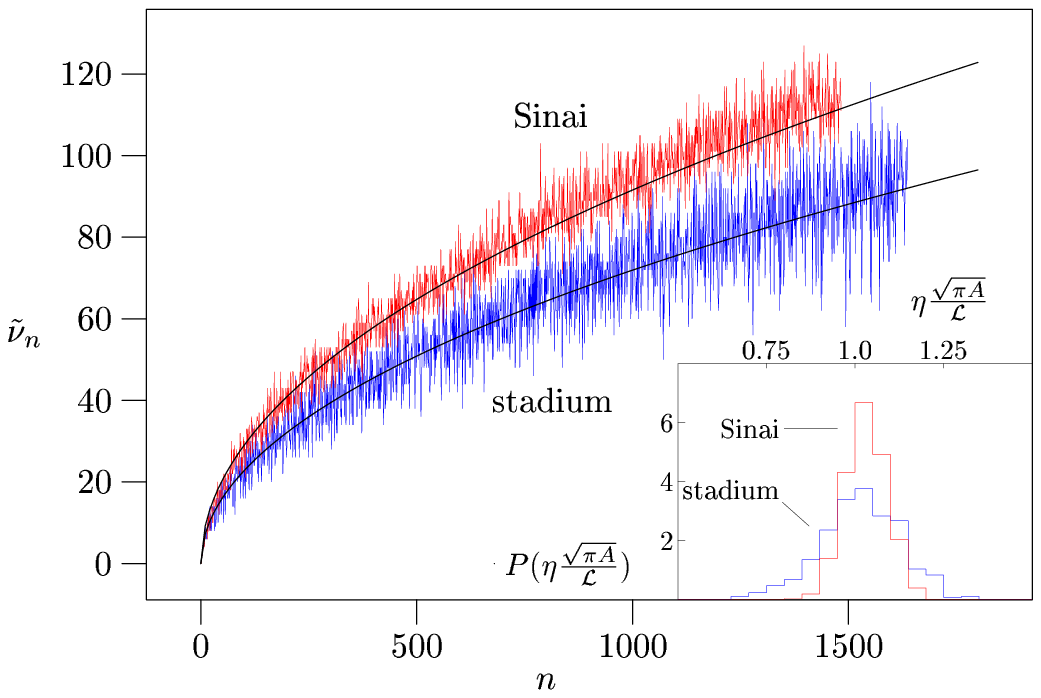}
    \caption{%
    The number of boundary intersections $\tilde \nu_n$ as a function of
$n$ for
the stadium and
the Sinai billiards. Smooth line - the   random
waves result(\ref {eq:meanvar}).  Inset - the $\eta  \frac{\sqrt{
\pi A}}{\mathcal L}$ distribution.  }
\label{fig:billbound}%
%\end{center}%
\end{figure*}

\noindent ({\it i.})   The distribution of boundary intersections. Given a
function
of the type (\ref {eq:wavefun}) we computed the number of the zeros of its
normal
derivative on a circle of radius  $R$ ($L=kR$). Denoting the normal derivative by
$u(\theta) = \partial_r \Psi(r,\theta)|_{r=R}$ the number of its zeros is \cite
{Kac}
\begin{equation}
\tilde \nu_{u} =\int_{0}^{2\pi}{\rm d}\theta\int_{-\infty}^{\infty}
\int_{-\infty}^{\infty}  {{\rm d}\xi{\rm d} \eta\over2(\pi \eta)^2}{\rm
e}^{i \xi u(\theta)
}(1-{\rm e}^{i
\eta
 \dot u(\theta)})
\label {eq:nzero}
\end{equation}
where $\dot u(\theta)={\rm d}_{\theta}u(\theta)$.
The mean number of zeros $\langle \tilde \nu \rangle$, and its variance  were
evaluated by performing the Gaussian integrals, with the result
\begin{eqnarray}
\label{eq:meanvar}
\langle \tilde \nu \rangle &=& kR \ \sqrt { 1+({2\over kR})^2}\
\approx  kR \nonumber \approx {k{\mathcal L}\over 2\pi} \\
var(\tilde \nu) &\approx & 0.0769\ k{\mathcal L}  \ .
\end{eqnarray}
The expression for $\langle \tilde \nu \rangle$ is exact, while that for the
variance is the leading term in the semiclassical limit. The variance was
computed
analytically up to an integral whose numerical value is
quoted.  Since  for a billiard of area $A$,  $n=\frac{Ak^2}{4\pi}$, (\ref
{eq:meanvar}) supports the proposed scaling (\ref {eq:nnbound}). Moreover, it
implies that the distribution of the scaled  number of  nodal intersections
limits
to a delta function centered at $ \eta  \frac{\sqrt{
\pi A}}{{\mathcal L}}  =1$. The agreement between the billiards data shown
in Fig. (\ref
{fig:billbound}) and the predictions of the random waves model, lends very
strong
support to our conjecture. However, the fluctuations in
$\nu_n$ are  affected by the presence of ``bouncing ball" states, which
explains the apparent
difference between the widths of the
$\eta$ distributions for the two billiards.
%%%%
%%%%
\noindent  ({\it ii.}) Signed area distribution.
Denote by  $|\Omega(\Psi)|_{\pm} $ the total area where the wave function
is positive
(negative). Clearly $\langle |\Omega|_{\pm} \rangle =|\Omega|/2$, where
$|\Omega|$ is the billiard area. To compute the signed area variance,
$\langle \left
(|\Omega|_{+}-|\Omega|_{-}\right)^2\rangle/|\Omega|^2 $, we used  the identity
\begin{equation}
|\Omega(\Psi)|_{\pm} = \frac{1}{2\pi i} \lim_{\epsilon\rightarrow
0^+}\int_{\Omega} {\rm d}^2r
\int_{-\infty}^{\infty}{\rm d}\xi \frac { {\rm e}^{\pm i \xi\ \Psi({\bf
r})}}{\xi-i\epsilon}
\ .
\label{eq:omega}
\end{equation}
%%%%%%%
 \begin{figure*} %
 % \begin{center}%
%\vspace{2.5cm}
\hspace{ -.25cm}
    \includegraphics[width=.9\linewidth]
    {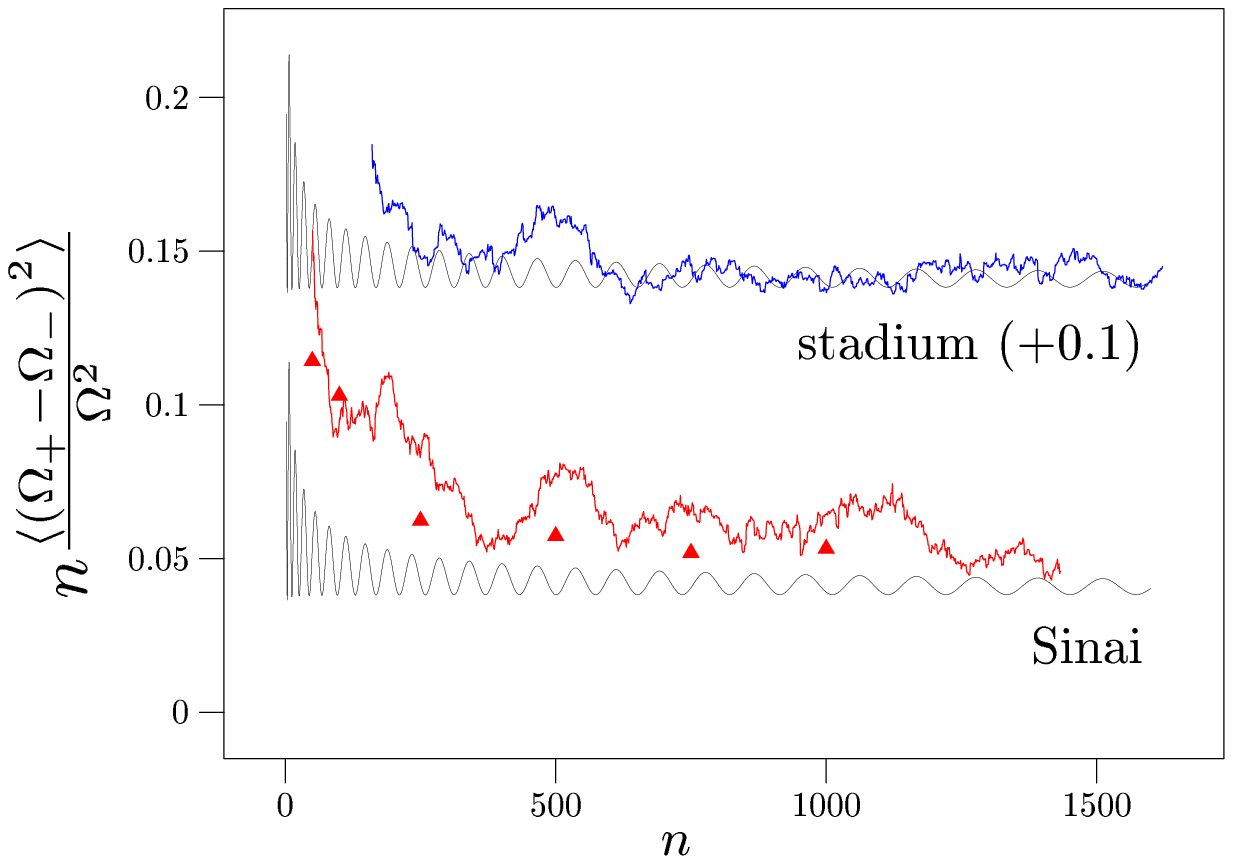}
    \caption{%
    The normalized signed area variance for the stadium and Sinai
billiards. Smooth curve -
analytic expression for random waves.
 Triangles -  numerical simulation for random waves (see text). }
\label{fig:billbound1}%
%\end{center}%
\end{figure*}%
%%%%
For a circle of radius $R$ the signed area variance can be reduced to an
expression involving
a single integral which was evaluateded numerically, and is shown in Fig.
4. as a function of
$n=(R k/2)^2$. For large $n$,
 \begin{equation}
{1\over  |\Omega|^2} \langle \left
(|\Omega|_{+}-|\Omega|_{-}\right)^2\rangle \approx  0.0386 n^{-1}\ .
\label{eq:areas}
\end{equation}
Fig. 4. shows also the data for the stadium and the  Sinai billiards,
which converge to the asymptotic limit (\ref {eq:areas}). The slow
convergence of the Sinai data is due to the exsitence of corners with sharp
angles
in the desymetrized Sinai billiard. The effect of such corners on the wave
function is
accenuated at low energies, and we checked this effect, by computing the
variance in a random
wave model for which the only angular waves allowed are multiples of
$4$ (triangular dots). This simulates the boundary conditions on the
symmetry line of the
Sinai billiard, and shows similar slow convergence to the asymptotic value.

%%%%%%%
 \begin{figure*} %
 % \begin{center}%
%\vspace{3.5cm}
\hspace{ -.4cm}
\includegraphics[width=.9\linewidth]{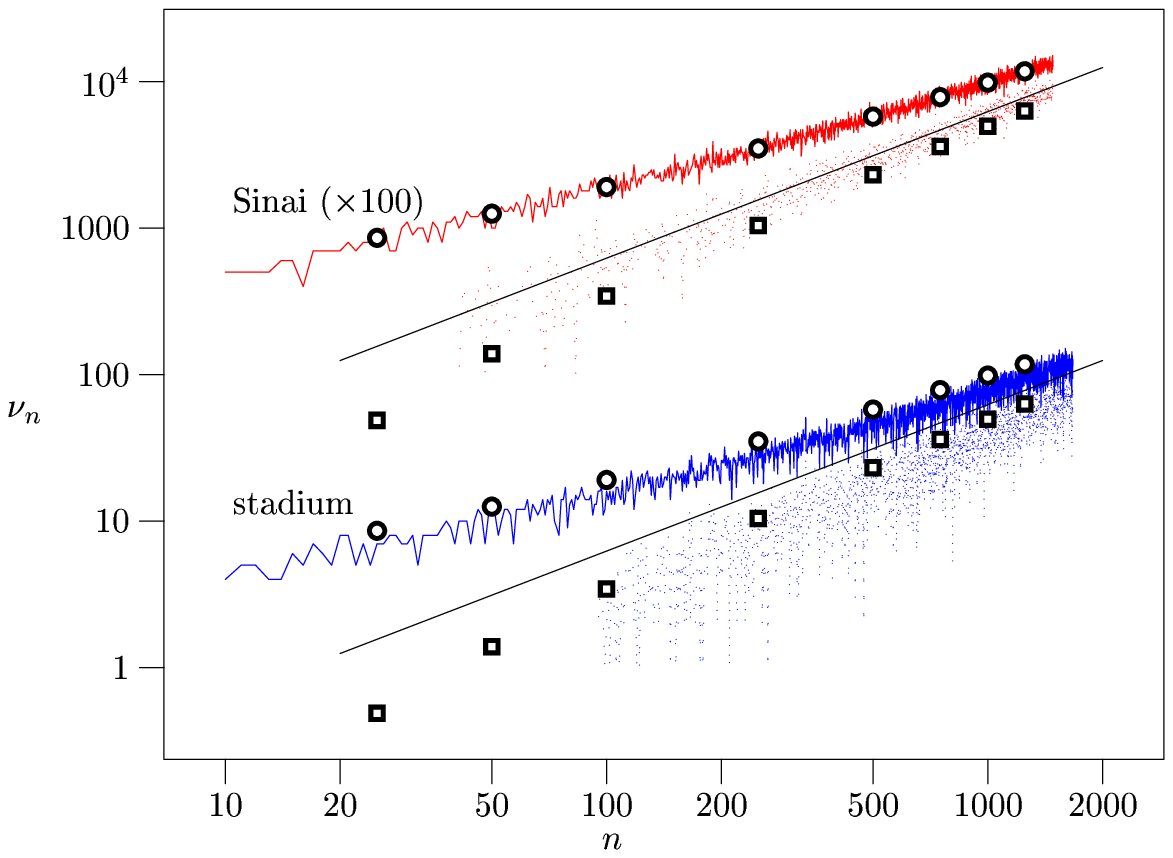}
\caption{
  The  number of nodal domains $\nu_n$ (solid lines) and inner
  nodal domains (dotted) for the Sinai and the stadium
  billiards, and for
  random waves (circles for nodal domains and squares for interior nodal
  domains). Smooth  lines:
  the prediction of the theory of Bogomolny and Schmit. 
  }
\label{fig:billbound2}%
                                %\end{center}%
\end{figure*}
%%%%
\noindent ({\it iii.})  The number of nodal domains $\nu_n$. The number of
nodal domains for the stadium and Sinai billiards are plotted in Fig. (5).
The predicted
$\langle \nu_n \rangle$ for the random waves model are shown as large
circular dots. These
data were computed numerically by sampling the random wave ensemble,
assuming a square
domain. The agreement  between the billiards and the random waves results
is satisfactory,
and confirms the expected scaling $\nu_n\approx n$ for high $n$. The random
and the billiards
data sets, however, show clear deviations from the expected asymptotic scaling
for low $n$ values, where  $\nu_n \approx n^{1/2}$.   To check the reason
for this behaviour
we computed and plotted the number of {\it interior} nodal domains - those
domains whose
boundaries do not include the billiard boundaries.  Their number for the
billiards and
the random waves model scale as $n$, suggesting the following explanation:
At low $n$, most
of the domains are boundary domains whose number scales as the number of
boundary
intersections, that is, $n^{1/2}$. Only when the wavelengths are
sufficiently smaller than the
typical linear dimension of the billiard, the majority  of domains are
interior, and their
number increases linearily with $n$.  The smooth lines in Fig 5.  represent
the
analytical result of \cite{bogoschmidt} for  $\langle \nu_n \rangle $.
This theory does not
 include the effects of the boundaries, and therefore it applies in the
limit of high $n$. It
is consistent with our numerical data, and confirms the linear scaling with
$n$.
The numerical simulations based on the random waves model, supported by the
results of \cite{bogoschmidt}  predict that $var(\nu_n)$ scales linearily with $n$ for
large $n$. Thus, the
limiting distribution $P(\xi)$ for chaotic billiards is expected to limit
to a delta
function, centered at the value predicted by \cite {bogoschmidt} to be $
0.0624$, much less than the Pleijel bound, or the values observed for the
square and circle
billiards.

 To summarize, we provided here ample evidence to show that the study of
nodal domains
counts provides a clear criterion of quantum chaos, which is not directly
related to
spectral statistics. Our results add a new dimension to previous attempts
to attribute
the morphology of wave functions to the nature of the underlying classical
dynamics
\cite{Shnirelman,Gutzwiller,Berry77,Heller,Aurich,Simmel}.

   This work was supported by the Minerva Center for non-linear
Physics at the Weizmann Institute. S.G. acknowledges support from the
Minerva Foundation.  We
thank E. Bogomolny and C. Schmit for allowing the use of their results
prior to publication,
and to Y. Avron, Z. Rudnick, Y. Sinai, M. Solomiak and D. Volk for
stimulating discussions and
comments.

\bibliographystyle{prsty}

\end{document}